\documentclass[prb,twocolumn,showpacs,preprintnumbers,amsmath,amssymb]{revtex4}

\usepackage{graphicx}
\usepackage{dcolumn}
\usepackage{bm}
\usepackage{amssymb}
\usepackage{graphicx}
\usepackage{amsmath}
\usepackage{xspace}

\begin{document}

\title{Ferromagnetic spinel CuCr$_2$Se$_4$ studied by Raman spectroscopy and
    lattice dynamics calculations}

\author{V. G. Ivanov$^1$, M. N. Iliev$^2$, Y.-H. A. Wang$^3$ and A.~Gupta$^3$}
\affiliation
{$^1$ Faculty of Physics, University of Sofia, 1164 Sofia, Bulgaria\\
$^2$Texas Center for Superconductivity and Department
of Physics, University of Houston, Texas 77204-5002, USA\\
$^3$Center for Materials for Information Technology and Department
of Chemistry, University of Alabama, Tuscaloosa, Alabama 35487,
USA}

\date{\today}

\begin{abstract}
The lattice dynamics of the ferromagnetic spinel CuCr$_2$Se$_4$ (T$_{\rm C} = 430$~K)
was studied experimentally by measuring the Raman spectra and theoretically by
calculation of zone center phonon frequencies within a shell model. All
Raman allowed modes ($A_{1g} + E_g + 3F_{2g}$) were identified and assigned
to specific  atomic motions. The relative intensity of the Raman lines varies strongly
with excitation photon energy between 1.58 and 2.71~eV, but no significant phonon
anomalies are observed near T$_{\rm C}$.
\end{abstract}
\maketitle

\section{Introduction}
In the recent years there is a revived interest in the properties of ferromagnetic chromium chalcogenide
spinels MCr$_2$X$_4$ (M=divalent element, X=S,Se,Te). These materils are prospective candidates for spin-based
electronic (spintronic) applications as the strong interaction between the
electronic and spin subsystems results in drastic changes  of the electronic transport and optical properties near the Curie temperature, T$_{\rm C}$.\cite{wojtowicz1969,haas1973,ramirez1997} CuCr$_2$Se$_4$
($T_C = 430$~K) attracts particular attention as it is ferromagnetic at room temperature and some theoretical
models predict that with suitable doping it may become half-metallic -
being an excellent metal for one spin channel and  excellent insulator for
the other spin channel.\cite{wang2009} While the structural, magnetic and electric properties of CuCr$_2$Se$_4$
have been subject of numerous studies, there are to our knowledge no reports on its lattice dynamics and possible
effects of spin-phonon coupling. In this work we present and analyze the experimental Raman spectra of CuCr$_2$Se$_4$ single crystals
in close comparison with predictions of lattice dynamics calculations within a shell modes.
 The Raman mode intensities exhibit resonant behavior for excitation
photon energies between 1.58 and 2.71~eV, but no significant anomalies in phonon parameters are
observed near the magnetic transition at $T_C$.

\section{Samples and Methods}
Single crystals of CuCr$_2$Se$_4$ were grown using the Chemical Vapor Transport (CVT)
technique following a procedure similar to that described in Ref.\onlinecite{masumoto1976}.
Briefly, polycrystalline CuCr$_2$Se$_4$ was used as the precursor with iodine gas as transport
agent placed in a sealed quartz tube (length: 15~cm; diameter: 1.5~cm).
The source temperature was maintained at 870$^\circ$C while the crystal growth at the other end of
the tube occurred at 770$^\circ$C. The reaction was carried out for a period of 80~hours, after which
the tube was cooled down and broken to collect the single platelet-like crystals
with (111) surfaces (see Fig.1) deposited in the cooler zone.

The Raman spectra were measured with five laser lines (784, 633, 515, 488, and 458~nm) using a T64000
spectrometer equipped with microscope and heating stage. In temperature-dependent studies the laser
power at the probe spot (3-4~$\mu$m diameter) was kept below 0.1~mW to assure negligible local laser
heating.

\begin{figure}[htbp]
\includegraphics[width=5.0cm]{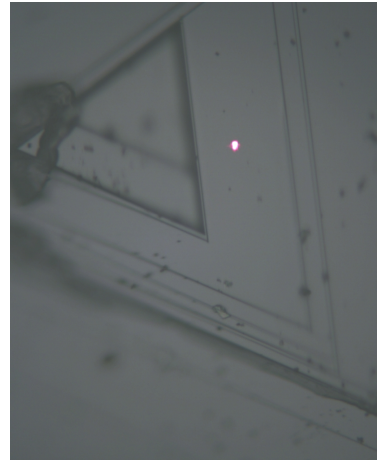}
\caption{(Color online) CuCr$_2$Se$_4$ surface topography. The laser focus spot is also seen.}
\end{figure}
\section{Results and Discussion}
The structure of MCr$_2$X$_4$ is described by the $Fd\bar{3}m$ (No. 227) space group. The primitive cell contains
14 atoms and the reduction of the 42-dimensional representation $\Gamma$ at $\vec{k} = 0$ into irreducible representations gives \cite{bilbao}:
$$\Gamma = A_{1g} + E_g +  F_{1g} + 3F_{2g} + 2A_u + 2E_u + 5F_{1u} + 2F_{2u}$$
The $A_{1g}$, $E_g$, and the three $F_{2g}$ modes are Raman-active, Four of the five $F_{1u}$ modes are IR-active and one is an acoustic mode. The $F_{1g}$, 2$A_{2u}$, 2$E_u$, and 2$F_{2u}$ modes are silent. Simple calculations for backward scattering from a (111) surface (Table~I) show
that the scattered intensity should not depend on the rotation of the crystal surface around the propagation direction of the incident light, but  only on the angle $\beta$ between the incident $\vec{e_i}$ and scattered $\vec{e_s}$ polarizations.

\begin{table}
\caption{Polarization selection rules for backward scattering from a (111) surface.}

\

\begin{tabular}{|c|c|c|c|}
\hline
   &   & parallel & crossed \\
  Mode &   & $\vec{e_i}\parallel\vec{e_s}$ & $\vec{e_i}\perp\vec{e_s}$ \\
       & $ 0^\circ < \beta < 90^\circ$ & $\beta = 0^\circ$ & $\beta = 90^\circ$ \\
     \hline

  &   & & \\
$A_{1g}$ & $a^2\cos^2\beta$     & $a^2$ & 0 \\
$E_g$ &$b^2(\sin^2\beta + \cos^\beta)$  & $b^2$ & $b^2$ \\
$F_{2g}$ & $d^2(1-\frac{2}{3}\sin^2\beta)$ & $d^2$ & $\frac{2}{3}d^2$  \\
   &  & & \\
     \hline
\end{tabular}
\end{table}
\subsection{Raman spectra}
Fig.2 shows the Raman spectra of CuCr$_2$Se$_4$ obtained with 633~nm excitation at room temperature with
parallel($\vec{e_i}\parallel\vec{e_s}$) and crossed($\vec{e_i}\perp\vec{e_s}$) scattering configurations.
Based on the selection rules of Table~I
the symmetries of the Raman lines at 108~cm$^{-1}$ ($F_{2g}$), 144~cm$^{-1}$ ($E_{g}$),
170~cm$^{-1}$ ($F_{2g}$), 220~cm$^{-1}$ ($F_{2g}$), and
227~cm$^{-1}$ ($A_{1g}$), are unambiguously determined. The Raman mode frequencies of CuCr$_2$Se$_4$ are compared
to those of isostructural ZnCr$_2$Se$_4$\cite{wakamura1976}, CdCr$_2$Se$_4$\cite{iliev1978} and HgCr$_2$Se$_4$\cite{iliev1978a} in
Table~II.
\begin{figure}[htbp]
\includegraphics[width=7.5cm]{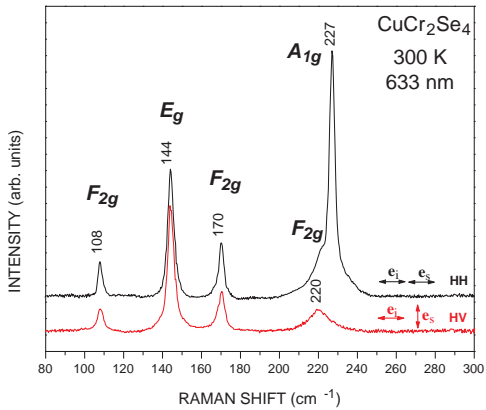}
\caption{(Color online) Raman spectra of CuCr$_2$Se$_4$ with parallel (HH) and crossed (HV) scattering configurations.}
\end{figure}

\begin{table}
\caption{Comparison of Raman mode frequencies (in cm$^{-1}$) of CuCr$_2$Se$_4$, ZnCr$_2$Se$_4$, CdCr$_2$Se$_4$, and HgCr$_2$Se$_4$ at 300~K.}

\

\begin{tabular}{|c|c|c|c|c|}
\hline
   &  &  & & \\
        & CuCr$_2$Se$_4$ & ZnCr$_2$Se$_4$ &CdCr$_2$Se$_4$ & HgCr$_2$Se$_4$\\
Mode    & this work      &Ref.\onlinecite{wakamura1976}  &Ref.\onlinecite{iliev1978} & Ref.\onlinecite{iliev1978a}\\
   &  &  & & \\
   \hline
      &  &  & & \\
 $A_{1g}$  & 227 & 240 &242 & 236\\
 $E_g$     & 144 & 152 &156 & 153\\
 $F_{2g}$  & 108 & 112 & 85 &  60\\
 $F_{2g}$  & 170 & 182 &169 & 164\\
 $F_{2g}$  & 220 & 230 &226 & 207\\
  &  &  & & \\
     \hline
\end{tabular}
\end{table}

The spectra obtained with 784~nm(1.58~eV), 633~nm(1.96~eV), 515~nm(2.41~eV), 488~nm(2.54~eV), and 458~nm(2.71~eV),
normalized to the intensity of the $E_g$ mode, are compared in Fig.3. The most obvious change
with increasing excitation photon energy is the increase of the relative intensity of the $F_{2g}$
modes near 170 and 220~cm$^{-1}$. Similar resonant behavior has been reported for CdCr$_2$Se$_4$.\cite{iliev1978}

\begin{figure}[htbp]
\includegraphics[width=7.5cm]{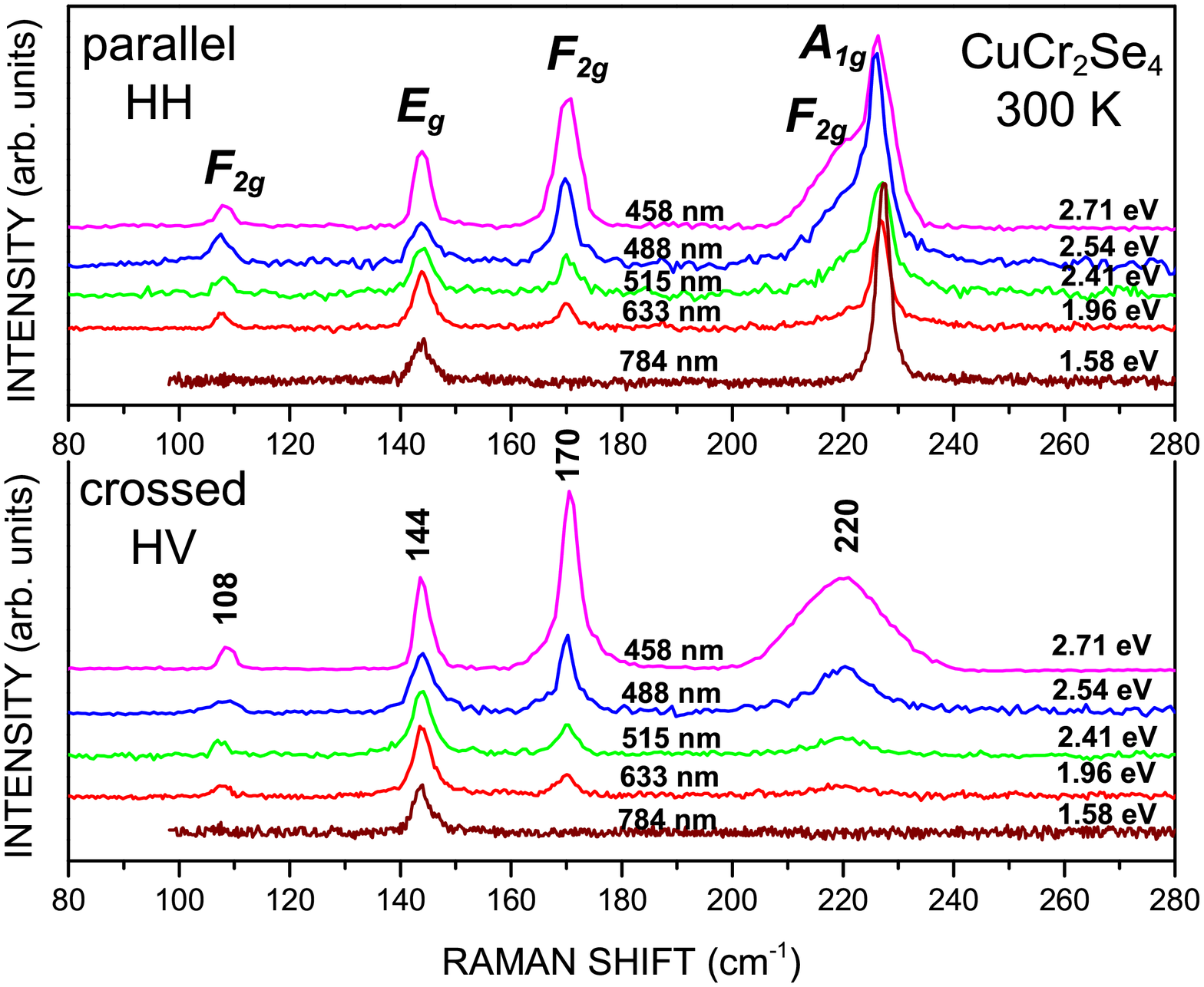}
\caption{(Color online) Raman spectra of CuCr$_2$Se$_4$ with 784~nm(1.58~eV), 633~nm(1.96~eV), 515~nm(2.41~eV),
488~nm(2.54~eV), and 458~nm(2.71~eV) excitation.}
\end{figure}

The temperature-dependent Raman spectra between 300 and 460~K as obtained with 633~nm excitation are shown in
Fig.4. The variations of the phonon parameters do not exhibit detectable anomalies near the magnetic transition that may be
considered as manifestation of strong spin-phonon coupling. Here again the results
are similar to those for CdCr$_2$Se$_4$ and HgCr$_2$Se$_4$.\cite{iliev1978,iliev1978a}

\begin{figure}[htbp]
\includegraphics[width=7.5cm]{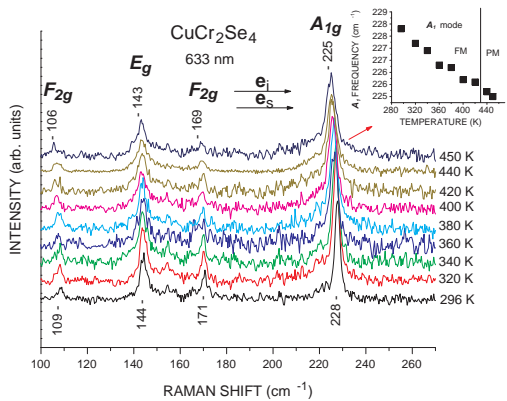}
\caption{(Color online) Temperature-dependent Raman spectra of CuCr$_2$Se$_4$ near the ferromagnetic transition (T$_C = 430$~K).
The inset shows the variations with T of the $A_{1g}$ frequency.}
\end{figure}

\subsection{Lattice Dynamics Calculations}
As it is clear from the experimental results, the polarized Raman measurements allow complete discrimination of the Raman-active modes of different symmetry. In the $E_g$ and $A_{1g}$ vibrations only Se atoms are involved.  The displacement vectors of the Se atoms in these modes are determined uniquely on the basis of symmetry considerations (see Table~III).  The corresponding displacement patterns are shown in Fig.5. The $A_{1g}$ mode is a symmetric radial breathing of the CuSe$_4$ tetrahedra while the $E_g$ mode is a result of Se displacement tangential to the Cu-Se bonds.   In $F_{2g}$ modes, however, a strong admixture of Cu and Se vibrations could be expected due to the comparable atomic masses of copper and selenium Thus, lattice dynamics calculations are necessary in order to obtain detailed information for the atomic displacements in these modes.

Lattice dynamical calculations for isostructural MCr$_2$X$_4$ chalcogenides (M=Zn,Cd,Hg; X=S,Se) reported before are based on valence force fields \cite{gupta1990,zwinscher1995} or a rigid-ion model (RIM).\cite{kushwaha2008} The valence force constants are of limited transferability in compounds with significant ionic bonding. As noted in Ref. \onlinecite{zwinscher1995}, the character of the various phonons as relating to eigenvectors and potential energy distribution differs to a much larger extent than expected for isostructural compounds. The rigid-ion model proposed in Ref. \onlinecite{kushwaha2008} is based upon pairwise cation-anion and anion-anion short range potentials. The potential parameters, however, are mapped onto a set of effective stretching and bending force constants constants. The reported effective force constants do not display a clear trend over the series of different M atoms, and their transferability in the case of M=Cu is not evident. Moreover, RIM does not account for the atomic polarizabilities and is unable to reproduce other important lattice properties as the high-frequency dielectric constant, the LO-TO splitting, as well as the off-diagonal elastic constants.

\begin{table}
\caption{Symmetry-adapted displacement vectors of the Se$(u,u,u)$ atom in the $A_{1g}$ and $E_g$ modes. The displacement vectors for the other atoms in the unit cell can be obtained by application of the corresponding symmetry transformation matrices.\cite{bilbao}}

\

\begin{tabular}{|c|c|}
\hline

Mode symmetry & Normalized displacement vectors\\
\hline
&  \\
$A_{1g}$ &      $\vec{e} = (1,1,1)/\sqrt3$\\
$E_g$   &   $\vec{e_1} = (2,-1,-1)/\sqrt6$\\
  &    $\vec{e_2} = (0,+1,-1)/\sqrt2$\\
\hline

\end{tabular}
\end{table}

\begin{figure}[htbp]
\includegraphics[width=6.5cm]{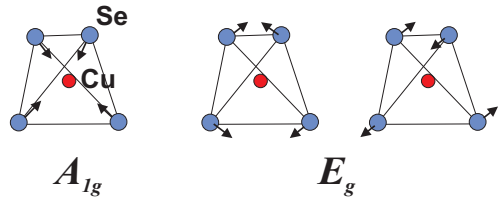}
\caption{(Color online) Tetrahedral displacements of the $A_{1g}$ and $E_g$ modes.}
\end{figure}

For the reasons outlined above we performed lattice dynamics calculations for CuCr$_2$Se$_4$ in the frame of the shell model (SM), which is the minimal extension over RIM capable to account for the ionic polarizability and to predict correct LO-TO splitting. In order to reduce the number of adjustable model parameters some physical simplifications were adopted. First, a valence shell was considered for the Se atoms only while Cu and Cr were treated as rigid ions. Second, the van der Waals attractive interaction was considered to act only between Se shells, while it was neglected for the Cu (Cr) core - Se shell pairs. These assumptions are justified by the much higher polarizability of the Se$^{-2}$ ion compared to that of the transition-metal ions. The rigid-ion approximation for transition-metals is a common approximation in the shell-model calculations on transition-metal oxides.\cite{Lewis} The short-range interatomic interactions were modelled by a Buckingham potential: $V(r) = A \exp(-r/\rho) - C/r^6$, where a non-zero van der Waals constant $C$ was retained for the Se shell - Se shell pairs only.

Even with the above simplifications the number of adjustable parameters exceeds the number of observables for CuCr$_2$Se$_4$, the lattice constant $a$, the Se fractional coordinate $u$ and the five Raman-mode frequencies. For this reason, the $A$, $\rho$ and $C$ parameters for the Se-Se interaction, the Se core and shell charges $X$ and $Y$, and the Se core-shell spring constant $k$  were adjusted initially by fitting of the lattice constant, static and optical dielectric constants, the LO and TO phonon frequencies, and the elastic modulus in ZnSe. As a next step these parameters were transferred to CuCr$_2$Se$_4$ and the parameters of the Cu-Se and Cr-Se core-shell interactions were optimized in order to fit the lattice parameter, the Se fractional coordinate $u$ and the frequencies of the Raman-active modes. The list of as obtained SM parameters is shown in Table~IV.

\begin{table}
\caption{Adjusted shell-model parameters for CuCr$_2$Se$_4$.}

\

\begin{tabular}{|c|c|c|c|}
\hline

 & & & Core-shell \\

Atom&Core charge&Shell charge&spring constant\\
     &  $X$ &$Y$ &$k$ (eV/{\AA}$^2$)\\
\hline
 & & & \\
Cu       &       1.00            &                    &                                         \\
Cr       &       2.30            &                    &                                         \\
Se       &       2.19        & -3.59              &69.7\\
 & & & \\
\hline
 & & & \\
Atomic pair     & $A$ (eV)&     $\rho$ (\AA) & $C$ (eV$\times${\AA}$^6$)\\

\hline
 & & & \\
Cu core - Se shell &    6827.4 & 0.2364 & 0 \\
Cr core - Se shell & 13817.5 & 0.2513 & 0 \\
Se shell - Se shell & 1482.2 & 0.3440   & 136.3 \\
 & & & \\
 \hline
\end{tabular}
\end{table}

\begin{figure}[htbp]
\includegraphics[width=8.5cm]{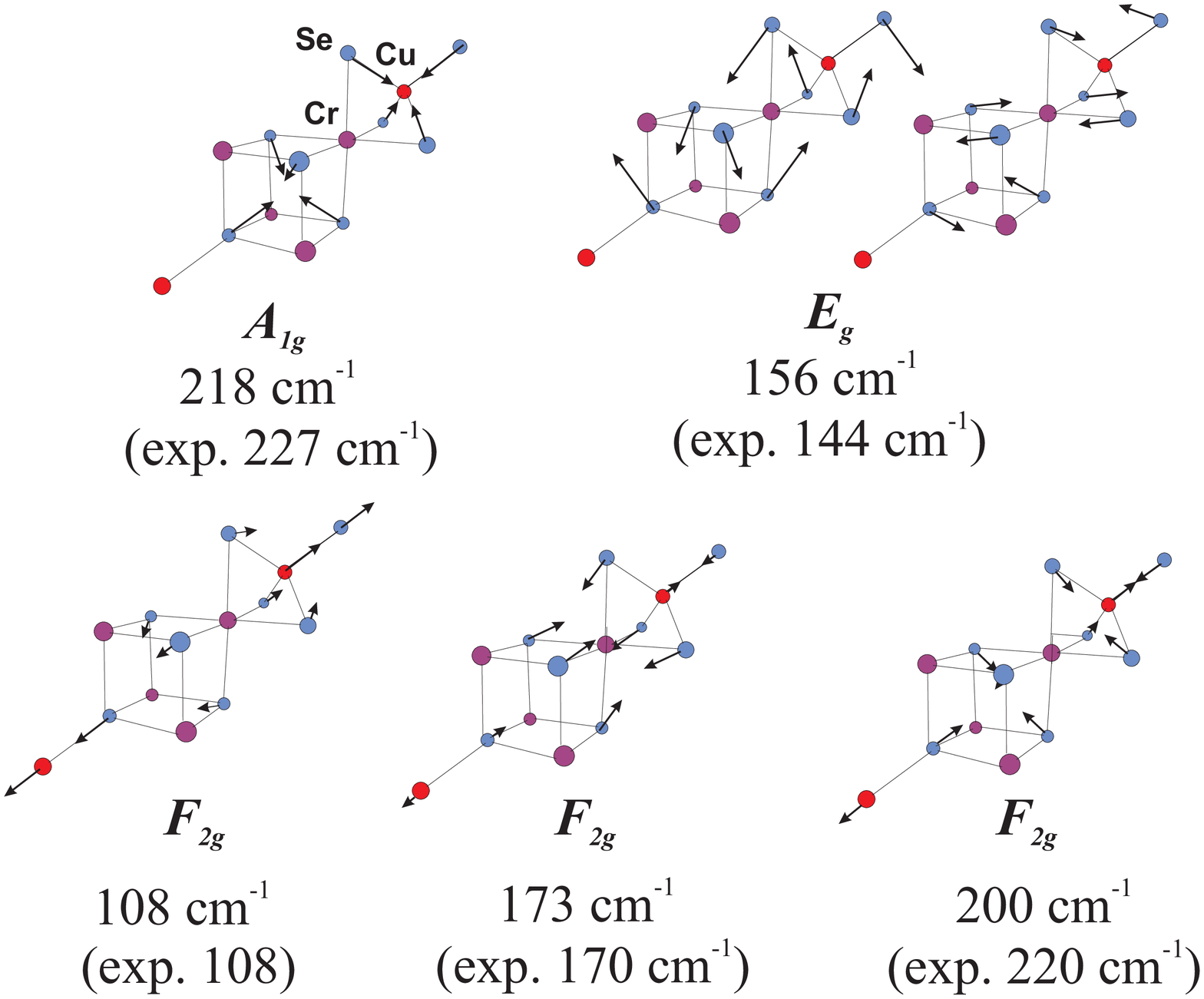}
\caption{(Color online) Atomic displacement of the Raman modes of CuCr$_2$Se$_4$.}
\end{figure}

A comparison between the calculated and the experimental lattice parameters and Raman-mode frequencies is shown in Table~V. The calculated atomic displacements for the three $F_{2g}$ modes are depicted in Fig.6. The lowest-frequency mode at 108~cm$^{-1}$ consists of an almost pure Cu translation within a practically static Se cage. This assignment corroborates the fact that the mode frequency scales with the mass $M$ of the A atom (A = Cu, Cd, Hg) as $M^{-1/2}$ (see Table-II). As an additional piece of evidence, we performed phonon calculations by substituting Cu with Cd and Hg and keeping the same values for the interatomic potentials. We obtained the frequencies of the lowest $F_{2g}$ mode at 83~cm$^{-1}$ for A = Cd and 63~cm$^{-1}$ for A = Hg in perfect agreement with the experimental values. The medium-frequency mode at 173~cm$^{-1}$ can be described as Cu translation against the Se cage, which moves in the opposite direction. The highest-frequency vibration at 200~cm$^{-1}$ is a superposition of an in-phase stretching of the Cu-Se bonds and Se-Cu-Se angle bending.

\begin{table}
\caption{Cell parameters, frequencies (in cm$^{-1}$) and atomic motions.}

\

\begin{tabular}{|c|c|c|}
\hline

Cell parameters &Exp & Calc \\
\hline
Lattice constant, $a$ ({\AA})  &10.337 & 10.363 \\
 Se position, $u$  & 0.382& 0.383 \\
    & &  \\
 \hline
 \end{tabular}

 \begin{tabular}{|c|c|c|c|}
 \hline
  & \multicolumn{2}{|c|}{Frequency}& Mode description\\
 Mode  &Exp &Calc  & in terms of CuSe$_4$ vibrations\\
   \hline
 $F_{2g}$ &108 &108  & Cu translation\\
  $E_g$  &144 & 156 & Se-Cu-Se bending\\
  $F_{2g}$ &170 & 173  & Cu translation agains Se\\
   $F_{2g}$ &220 & 200 & Cu-Se stretching + Se-Cu-Se bending\\
   $A_{1g}$ & 227& 218 & symmetric Cu-Se stretching\\
   \hline

  \end{tabular}
   \end{table}

   \

\subsection{Resonant Behavior of Raman Scattering}

 The Raman scattering cross-section $S(\hbar\omega)$ in non-transparent materials is related to the combined electronic density of states. As a rule $S(\hbar\omega)$ is proportional to the $|\frac{d \epsilon(\hbar\omega)}{d \hbar\omega}|^2$ and has maxima near interband gaps. The experimentally observed dependence of the Raman intensity on the excitation photon energy $I(\hbar\omega)$, however, is additionally modified by the spectral properties of the Raman setup and the absorption and reflection losses, which occur for the incident and scattered light. The detail analysis of resonant Raman scattering in closely related CdCr$_2$Se$_4$ has shown that $S(\hbar\omega)$ has a maximum near 2~eV.\cite{iliev1978} It has also been shown in that study that the intensity of the $F_{2g}$ Raman lines, if normalized to the $E_g$ intensity, strongly increases with photon energy between 1.5 and 2.8~eV thus demonstrating the dependence of scattering cross section on the phonon symmetry. As it follows from Fig.3,  similar enhancement of the relative intensity of the $F_{2g}$ modes is observed for CuCr$_2$Se$_4$. Fig.7 shows in more detail the variation with $\hbar\omega)$ of the normalized intensity for the 170~cm$^{-1}(F_{2g})$, 220~cm$^{-1}(F_{2g})$, and 227~cm$^{-1}(A_{1g})$ modes.

For tentative explanation of the resonant enhancement of the $F_{2g}$ intensity let us consider the {\it ab initio} electronic structure calculations \cite{Ogata,Antonov} for CuCr$_2$Se$_4$. Following these calculation the electronic density of states (DOS) above the Fermi level is dominated by the Cr $d$ states, while those below the Fermi level by the Cu $d$ and the Se $p$ states.  An optical transition between Cu and Cr $d$ bands can be disregarded due to the large spatial separation between the Cu and Cr ions. Therefore, a possible candidate for an optical transition could be an electron transfer between the Se $p$ and Cr $d$ orbitals. Transition energies of 3.2~eV between Se $p_{1/2}$ and Cr $d_{3/2}$ states, and 2.84~eV between Se $p_{3/2}$ and Cr $d_{5/2}$ states could be expected on the basis of calculated centers of gravity of the corresponding bands \cite{Antonov}. The latter value is close the experimentally observed maximum at 2.75~eV of the optical conductivity of CuCr$_2$Se$_4$ reported by Ongushi et al.\cite{ongushi2008}

\begin{figure}[htbp]
\includegraphics[width=5cm]{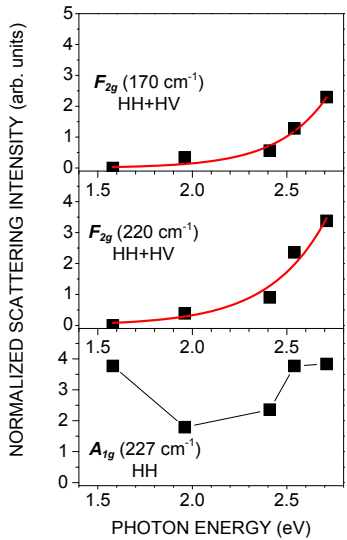}
\caption{(Color online) Variation of the normalized Raman intensity of the 170~cm$^{-1}(F_{2g})$, 220~cm$^{-1}(F_{2g})$, and 227~cm$^{-1}(A_{1g})$ modes with excitation photon energy.}
\end{figure}

The lattice dynamics calculation can help us understand why A$_{1g}$ and $F_{2g}$ modes exhibit stronger coupling  to Se~$p$ - Cr~$d$ transitions than the $E_g$. The coupling of a Raman-active phonon to an optical transition is mainly due to two mechanisms: (i) modulation of the dipolar transition moment, and (ii) modulation of the transition energy by the atomic displacements. As a rule the modulation of the transition energy has stronger resonant behavior and gives a dominant contribution to the Raman scattering intensity under resonant conditions. We will consider the optical $p \rightarrow d$ transitions in a simplified tight-binding picture assuming that the transition energy is roughly equal to the difference between the on-site energies of the Se $p$ and Cr $d$ orbitals. Since the on-site energy is a scalar quantity, it transforms according to the permutational representation of the given atomic position. The decomposition of the permutational representations for the 16d (Cr) and 32e (Se) positions gives:
\begin{eqnarray}
\Gamma(\text{Se})=A_{1g}+A_{2u}+F_{2g}+F_{1u}\\
\Gamma(\text{Cr})=A_{1g}+F_{2g}
\end{eqnarray}
Therefore, among Raman-active modes only A$_{1g}$ and F$_{2g}$ modulate the Se  and Cr orbital energies and respectively the energy of the $p \rightarrow d$ transitions. This fact can be understood by inspecting the atomic displacement patterns in Fig. 6. In the A$_{1g}$ and F$_{2g}$ modes the Se displacements have identical radial components relative to the Cr atom and, therefore, give identical variations of the Madelung potential at the Cr site. For the $E_g$ mode, however, the radial components of the Se displacements are of alternating signs leading to cancelation of the Madelung potential variations at the Cr site. By following our lattice dynamics calculations it is easy to note that the resonant enhancement of the F$_{2g}$ modes increases with the increase of the radial character of the vibration. Interestingly, the $A_{1g}$ mode shows an enhanced normalized intensity also with 784~nm(1.58~eV) excitation, which is in the energy range of the Cr $d-d$ transitions.\cite{ongushi2008}

\section{Conclusions}
The Raman spectra of the ferromagnetic spinel CuCr$_2$Se$_4$ ($T_C = 430$~K)
was studied experimentally and analyzed theoretically by calculation of lattice dynamics.
The normalized Raman scattering intensity for A$_{1g}$ and F$_{2g}$ exhibits resonant behavior similar to that known for
the isostructural CdCr$_2$Se$_4$, but no significant phonon anomalies are observed near $T_C$.

\acknowledgments
This work was support in part by the State of Texas through the
Texas Center for Superconductivity at the University of Houston
and by NSF MRSEC (Grant No.DMR-0213985) and partly by the contract \# DO 02-167/2008 of the Bulgarian National Fund for Scientific Research.

\end{document}